\documentclass[sigconf]{acmart}
\usepackage{graphicx}
\graphicspath{{./img/}}
\usepackage{array}
\newcolumntype{P}[1]{>{\centering\arraybackslash}p{#1}}
\newcolumntype{M}[1]{>{\centering\arraybackslash}m{#1}}

\usepackage{balance}

\copyrightyear{2022}
\acmYear{2022}
\setcopyright{acmcopyright}\acmConference[ICPP Workshops '22]{51th International Conference on Parallel Processing Workshop}{August 29-September 1, 2022}{Bordeaux, France}
\acmBooktitle{51th International Conference on Parallel Processing Workshop (ICPP Workshops '22), August 29-September 1, 2022, Bordeaux, France}
\acmPrice{15.00}
\acmDOI{10.1145/3547276.3548630}
\acmISBN{978-1-4503-9445-1/22/08}

\begin{document}

\title{Optimizing Hardware Resource Partitioning and Job Allocations on Modern GPUs under Power Caps}



\author{Eishi Arima}
\affiliation{%
  \institution{Technical University of Munich}
  \city{Garching}
  \country{Germany}
}
\email{eishi.arima@tum.de}

\author{Minjoon Kang}
\affiliation{%
  \institution{Technical University of Munich}
  \city{Garching}
  \country{Germany}
}
\email{minjoon.kang@tum.de}

\author{Issa Saba}
\affiliation{%
  \institution{Technical University of Munich}
  \city{Garching}
  \country{Germany}
}
\email{issasaba@hotmail.com}

\author{Josef Weidendorfer}
\affiliation{%
  \institution{Leibniz Supercomputing Centre}
  \city{Garching}
  \country{Germany}
}
\email{josef.weidendorfer@lrz.de}

\author{Carsten Trinitis}
\affiliation{%
  \institution{Technical University of Munich}
  \city{Garching}
  \country{Germany}
}
\email{carsten.trinitis@tum.de}

\author{Martin Schulz}
\affiliation{%
  \institution{Technical University of Munich}
  \city{Garching}
  \country{Germany}
  \\
  \institution{Leibniz Supercomputing Centre}
  \city{Garching}
  \country{Germany}
}
\email{martin.w.j.schulz@tum.de}


\begin{abstract}
CPU-GPU heterogeneous systems are now commonly used in HPC (High-Performance Computing). However, improving the utilization and energy-efficiency of such systems is still one of the most critical issues.
As one single program typically cannot fully utilize all resources within a node/chip, co-scheduling (or co-locating) multiple programs with complementary resource requirements is a promising solution. 
Meanwhile, as power consumption has become the first-class design constraint for HPC systems, such co-scheduling techniques should be well-tailored for power-constrained environments. 
To this end, the industry recently started supporting hardware-level resource partitioning features on modern GPUs for realizing efficient co-scheduling, which can operate with existing power capping features. 
For example, NVidia's MIG (Multi-Instance GPU) partitions one single GPU into multiple instances at the granularity of a GPC (Graphics Processing Cluster). 
In this paper, we explicitly target the combination of hardware-level GPU partitioning features and power capping for power-constrained HPC systems. 
We provide a systematic methodology to optimize the combination of chip partitioning, job allocations, as well as power capping based on our scalability/interference modeling while taking a variety of aspects into account, such as compute/memory intensity and utilization in heterogeneous computational resources (e.g., Tensor Cores). 
The experimental result indicates that our approach is successful in selecting a near optimal combination across multiple different workloads. 


\end{abstract}





\maketitle
\section{Introduction}\label{introduction}
Heterogeneous CPU-GPU architectures are now broadly used in a variety of computing systems. In particular, they are dominating in the HPC (High-Performance Computing) or supercomputing area. This kind of architecture first appeared and got ranked in the top 10 of the Top500 list about a decade ago, and now around 150 out of the 500 top-class supercomputers are GPU-accelerated systems (as of Nov 2021)~\cite{top500}. One reason behind this trend is that scientific applications generally expose much data parallelism and are throughput-limited. Thus, throughput-oriented architectures, like GPUs, are preferred. Further, recent GPUs generally have matrix-matrix multiplication engines (e.g., Tensor Cores~\cite{a100}), which is a useful feature for supercomputing centers where AI workloads occupy an increasingly large fraction of resources and runtime. 

Meanwhile, one of the most significant problems in supercomputers or any other large-scale systems, such as datacenters, is the resource waste inside a node. As computing nodes in such systems have become increasingly rich in terms of FLOPS, bandwidth, capacity, etc., it has become more difficult to fully utilize its resources by one single job. To combat this problem, the industry as well as the academia has introduced several concurrency management techniques to enable co-executing multiple jobs/applications/programs on one single GPU. 
As examples, Slate~\cite{slate} and smCompactor~\cite{smcompactor} are in-node process schedulers that provide automatic pair selections and dynamic optimizations. 
Another example is MPS (Multi-Process Service)~\cite{mps} that mixes multiple different CUDA programs to efficiently utilize SMs (Streaming Multiprocessors) inside of a GPC (Graphics Processing Cluster) in a time-sharing manner. 
Further, NVIDIA offers yet another sophisticated hardware-level resource partitioning feature, called MIG (Multi-Instance GPU), in their most recent products based on the Ampere architecture~\cite{a100}, which divides one single GPU into multiple parts at the granularity of GPCs.

Another significant issue in the supercomputing area is the enormous amount of power consumption, and developing sophisticated power management techniques will be essential, in particular in the post exa-scale era. 
In fact, the power consumption of top-class supercomputers or High-Performance Computing (HPC) systems have been increasing considerably over the past few decades. 
As a result, one of the most powerful supercomputers in the world now consumes a significant amount of power, almost hitting 30MW~\cite{top500}. 
Meanwhile, energy costs have been raising considerably and thus setting a power constraint on the entire HPC system in order to keep within a budgetary upper limit is becoming more and more critical. 
As a consequence, near-future HPC systems will be operated under a certain power constraint, 
and sophisticated power capping techniques that set power limits to jobs/components at different granularity for such environments are desired and thus are widely studied~\cite{gpu-pow-cap,cpu-gpu-cosh,lr-application4,lr-application5, overprovisioning,overprovisioning2}.

In this paper we explicitly target the combination of co-scheduling and power capping on modern GPUs, and provide a systematic approach to simultaneously optimize several knobs, i.e., resource partitioning, job allocations, and power capping, for co-located jobs. 
As an example, we utilize the MIG feature~\cite{mig}, the state-of-the-art concurrency control feature supported in the most recent GPUs~\cite{a100}, and we optimize the resource partitioning state, the job allocations to the partitions, as well as the chip-level power cap setup, depending on a given objective (or policy). 
To this end, we provide an optimization workflow that consists of linear regression performance modeling, model coefficients calibration, and decision making to select an optimal setup. 

The following are the major contributions of this paper: 
\begin{itemize}
\item As far as we know, \textit{this is the first work that attempts to optimize the resource partitioning, job allocations, and power budgeting at the same time on a real GPU chip using a hardware-level resource partitioning mechanism, i.e., the MIG feature}.  
\item To this end, we performed several preliminary observations on a MIG-enabled GPU and obtained the following insights: (1) the application scalability at the granularity of GPC highly depends on the memory sharing option, the power cap setup, and the application characteristics, including the usage of heterogeneous computation resources (e.g., Tensor Core utilization rate) and the memory/compute intensity; and (2) as a consequence, the best hardware setup and job allocations highly depend on these application characteristics. 
\item Based on these observations, we propose a systematic methodology to optimize the resource partitioning, the job allocations, as well as the chip power budgeting for a given problem (or policy), while using our linear regression performance model that takes the hardware setup and the application characteristics into account while consisiting of scalability and interference parts. 
\item Our experimental result clarifies that our approach is successful in selecting an optimal combination, achieving almost the best throughput/energy-efficiency across multiple different workloads for different objectives (or policies). 
\end{itemize}

The remainder of the paper is organized as follows. In the next section, we summarize the background of this study (e.g., our target systems, modern GPU architectures, and co-scheduling features). 
Then, Section~\ref{motivation} demonstrates our fundamental observations with respect to optimal hardware resource setups. 
We then provide the problem definition, the overview of our solution, and our linear regression modeling in Section~\ref{proposal}. 
Next, we specify our evaluation environment/methodology and the experimental result in Section~\ref{evaluation}. 
In Section~\ref{discussions}, we discuss the generality, applicability, limitations, and so forth. 
We then introduce a variety of related studies and clarify the difference from our work in Section~\ref{related}. 
Finally, we conclude this work in Section~\ref{conclusion}. 

\section{Background}\label{background}
In this section, we summarize the background knowledge related to this study. First, we explain our target scenario on co-scheduling and power management on HPC systems. Next, we introduce the MIG feature, the state-of-the-art resource partitioning feature supported in the most recent GPUs. 

\begin{figure}[t]
  \begin{center}
    \includegraphics[width=0.75\linewidth]{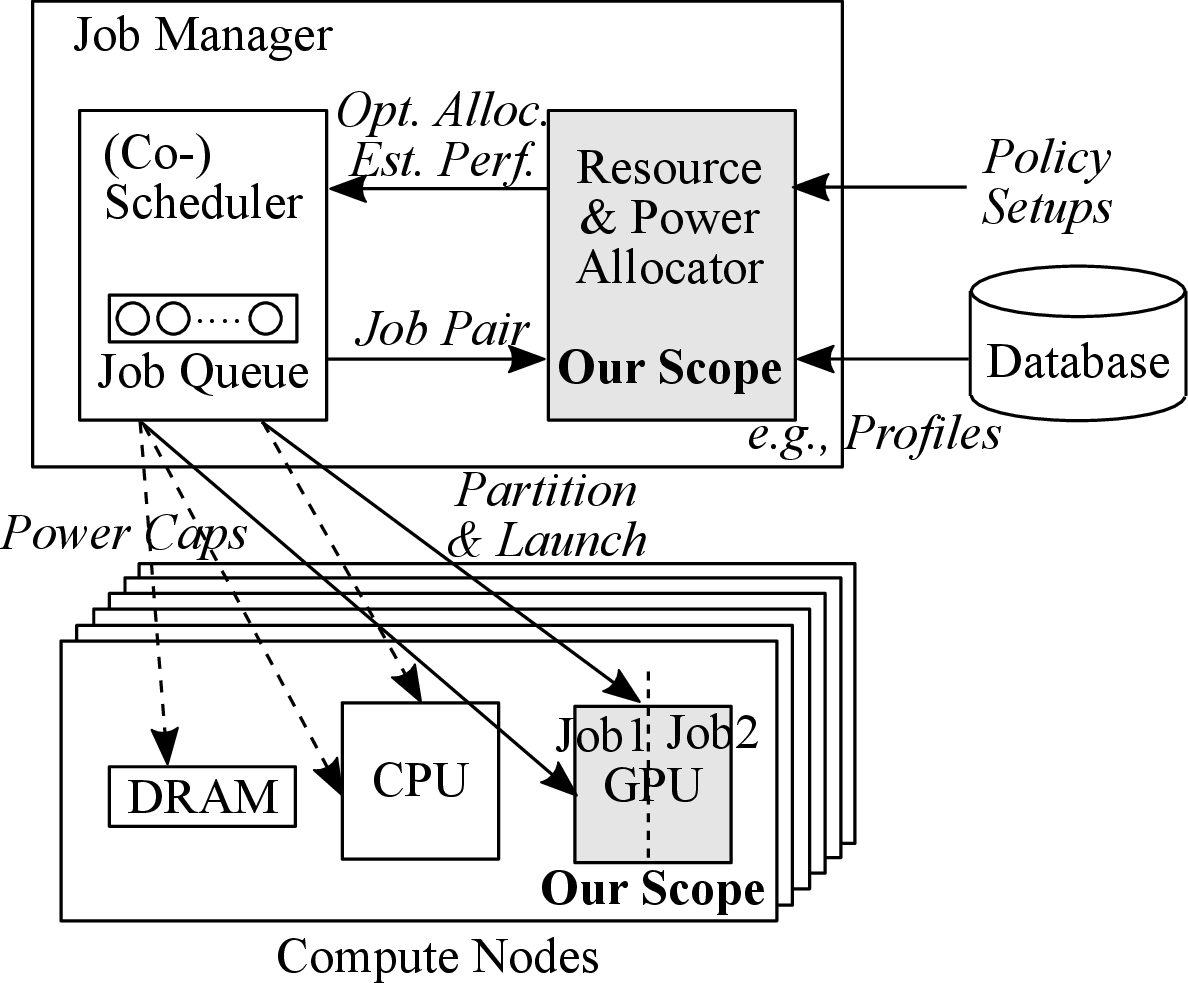}
    \vspace{-5pt}
    \caption{Our Assuming HPC System and Our Scope}
  \label{scheduling}
  \end{center}
  \vspace{-10pt}
\end{figure}

\subsection{Co-scheduling and Power Management on HPC Systems}

Figure~\ref{scheduling} illustrates the software/hardware architecture of our target HPC systems, consisting of multiple compute nodes. 
Each compute node is CPU-GPU heterogeneous, and the GPU supports a hardware-level resource partitioning feature such as MIG, which we will detail later. 
These compute nodes are controlled by a job management software that is capable of power management and co-scheduling, i.e., co-locating multiple jobs on the same node. 
More specifically, we assume the job manager sets power caps (or power limits) to several components (e.g., CPU, GPU, and DRAM) in each node/job and handles the resource partitioning and job mappings to them. 
As an example, thr SLURM workload manager~\cite{slurm}, a job scheduler widely used for HPC systems, supports the power capping feature via RAPL (Running Average Power Limit) interface~\cite{rapl} and is also capable of launching multiple applications on the same node.

In this study, we assume the job manager mainly consists of two parts: (1) \textit{(Co-)Scheduler} that manages the job queue, makes decisions on job launches/co-locations, and allocates power budgets to them; and (2) \textit{Resource and Power Allocator} that receives co-location job pair candidates from the Co-Scheduler and chooses/returns optimal resource/power allocations to each of them based on its performance modeling, which are ultimately used for the scheduling decisions. 
\textit{In this work, we focus on the latter part and target the resource partitioning, job mapping, and power budgeting on a GPU for a given job set.} 
For this, we define optimization problems (or policies), propose a profile-driven framework to solve the problem, and provide a simple performance model. 

Overall, our target scenario is becoming more and more realistic for near-future HPC systems. 
CPU-GPU heterogeneous architectures are now commonly used in HPC systems -- around 30\% of the 500 top-class supercomputers are GPU-accelerated systems (as of Nov 2021)~\cite{top500}. 
Due to the significant amount of power consumption and the increasing energy price, near-future HPC systems will be operated under a strict power constraint. As a consequence, sophisticated power management mechanisms that distribute power budgets across jobs/nodes/components depending on their need will be essential~\cite{gpu-pow-cap,cpu-gpu-cosh,lr-application4,lr-application5, overprovisioning,overprovisioning2}. 
Prior studies have shown that co-scheduling and resource partitioning increase the total system throughput and energy efficiency considerably~\cite{cpu-cosh-powcap,cpu-cosh-powcap2,cpu-gpu-cosh}, which should also be the case at the GPU chip level granularity.


\subsection{MIG: Hardware-Level Concurrency Support on Modern GPUs}

\begin{figure}[t]
  \begin{center}
    \includegraphics[width=0.9\linewidth]{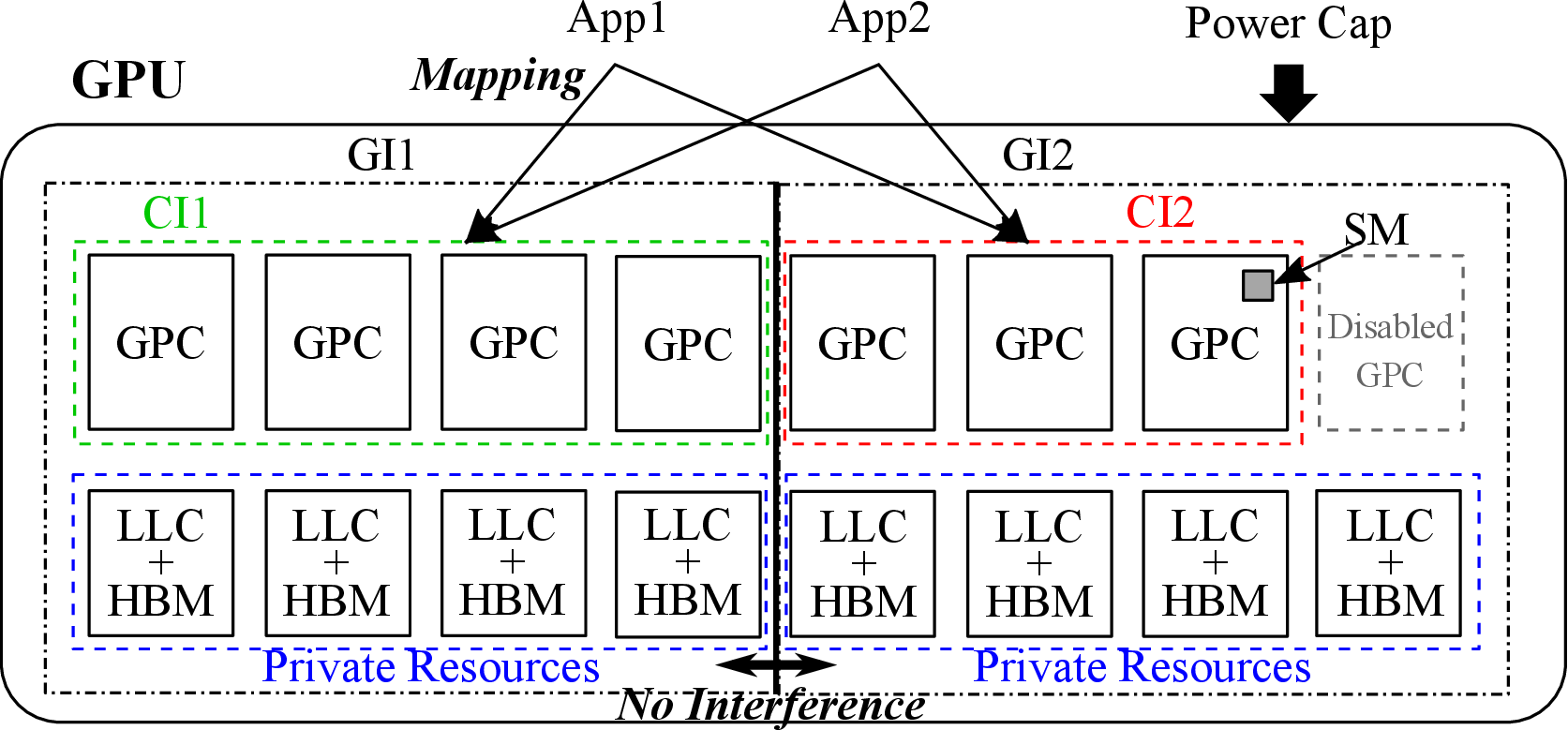}
    \vspace{-5pt}
    \caption{MIG with Private LLC/HBM Option}
  \label{mig-private}
  \end{center}

  \begin{center}
    \includegraphics[width=0.9\linewidth]{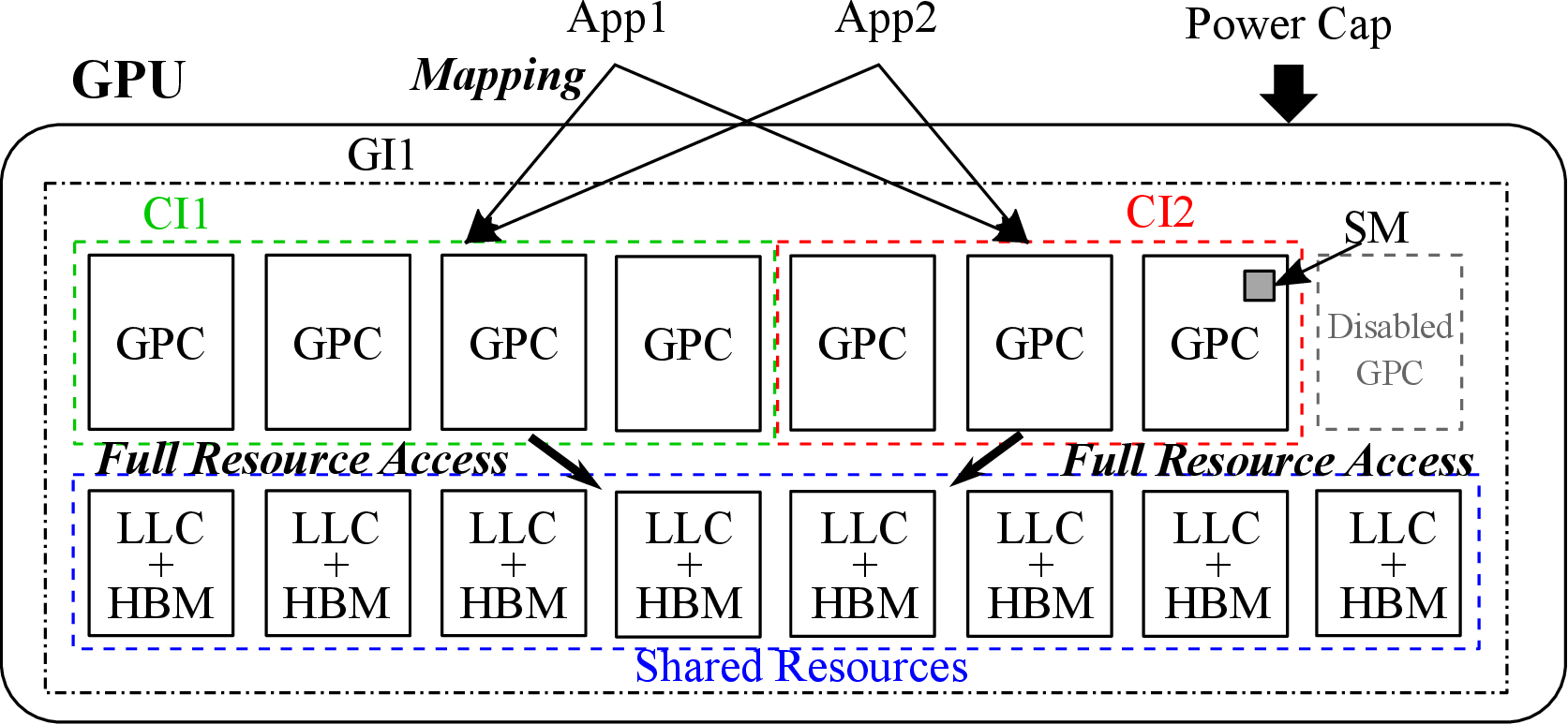}
    \vspace{-5pt}
    \caption{MIG with Shared LLC/HBM Option}
  \label{mig-shared}
  \end{center}
  \vspace{-5pt}
\end{figure}

\begin{figure*}[t]
{
\begin{tabular}{c}

 \begin{minipage}{\hsize}
  \begin{center}
  \begin{minipage}{0.22\hsize}
  \begin{center}
  \includegraphics[width=\linewidth]{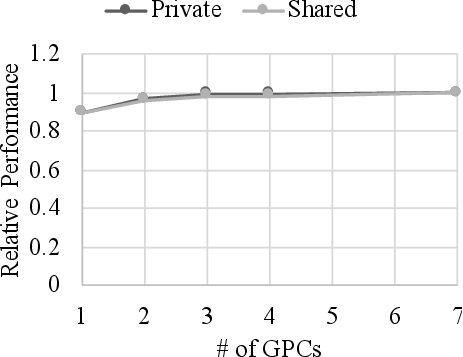}
  \textbf{kmeans}
  \end{center}
  \end{minipage}
  \hspace{0.01\hsize}
  \begin{minipage}{0.22\hsize}
  \begin{center}
  \includegraphics[width=\linewidth]{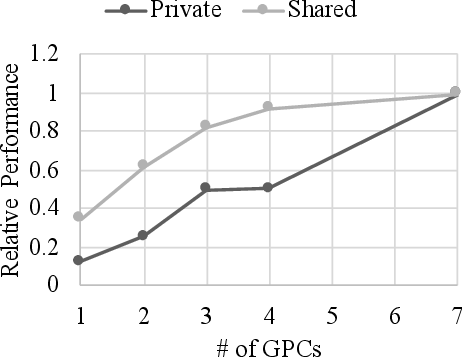}
  \textbf{stream}
  \end{center}
  \end{minipage}  
  \hspace{0.01\hsize}
  \begin{minipage}{0.22\hsize}
  \begin{center}
  \includegraphics[width=\linewidth]{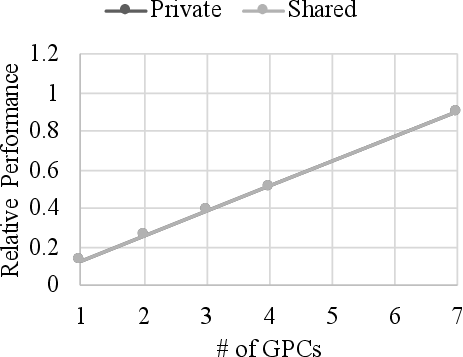}
  \textbf{dgemm}
  \end{center}
  \end{minipage}
  \hspace{0.01\hsize}
  \begin{minipage}{0.22\hsize}
  \begin{center}
  \includegraphics[width=\linewidth]{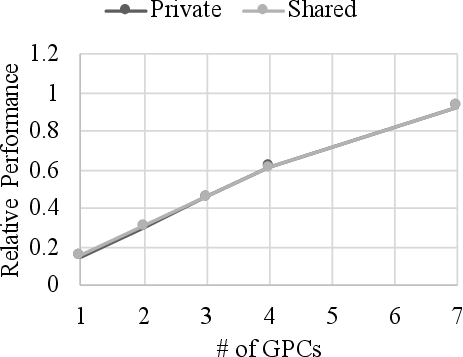}
  \textbf{hgemm}
  \end{center}
  \end{minipage}  
  \end{center}
 \end{minipage}
\end{tabular}
  \vspace{-5pt}
    \caption{Scalability Observations for Different Partitioning Options across Different Benchmarks (Power Cap: 250[W])}\label{scale-opt}
\vspace{10pt}
\begin{tabular}{c}
 \begin{minipage}{\hsize}
  \begin{center}
  \begin{minipage}{0.22\hsize}
  \begin{center}
  \includegraphics[width=\linewidth]{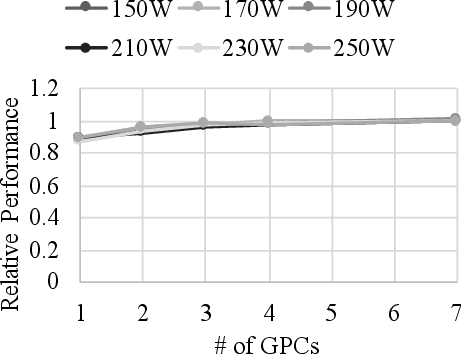}
  \textbf{kmeans}
  \end{center}
  \end{minipage}
  \hspace{0.01\hsize}
  \begin{minipage}{0.22\hsize}
  \begin{center}
  \includegraphics[width=\linewidth]{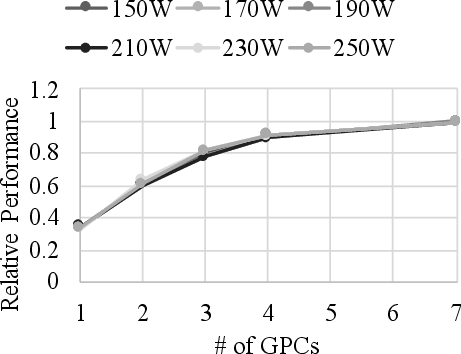}
  \textbf{stream}
  \end{center}
  \end{minipage}  
  \hspace{0.01\hsize}
  \begin{minipage}{0.22\hsize}
  \begin{center}
  \includegraphics[width=\linewidth]{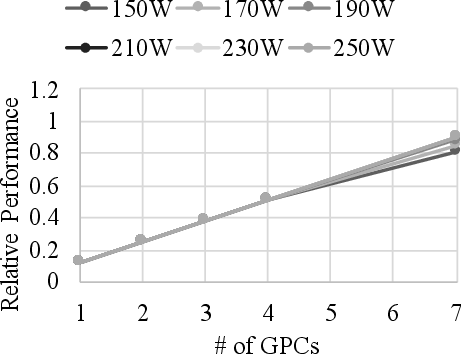}
  \textbf{dgemm}
  \end{center}
  \end{minipage}
  \hspace{0.01\hsize}
  \begin{minipage}{0.22\hsize}
  \begin{center}
  \includegraphics[width=\linewidth]{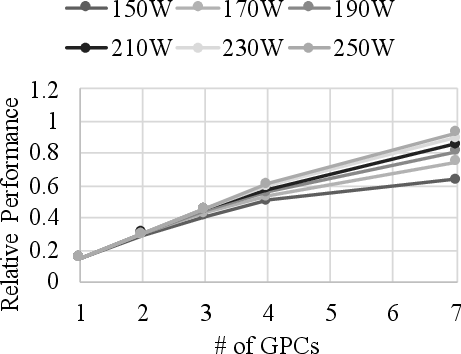}
  \textbf{hgemm}
  \end{center}
  \end{minipage}  
  \end{center}
 \end{minipage}
\end{tabular}
  \vspace{-5pt}
    \caption{Scalability Observations for Different Power Cap Setups across Different Benchmarks (Partitioning: Shared)}\label{scale-pow}
}
\end{figure*}

To this end, we target GPUs that support both a GPU partitioning and a power capping feature. As an example, current state-of-the-art commercial GPUs, i.e., NVIDIA's Ampere generation GPUs, such as the A100~\cite{a100}, support Multi-Instance GPU (MIG) that partitions a GPU into multiple divisions at the hardware level~\cite{mig}. 
Figure~\ref{mig-private}/\ref{mig-shared} illustrates how the MIG feature operates with different partitioning options. 
The MIG divides a GPU in a hierarchical manner: it first sets up GPU Instance(s) or GI(s) and then launches one or more Compute Instance(s) or CI(s) inside of a GI. 
A unique UUID\footnote{Universally Unique IDentifier} is allocated to each of the CI (and thus these CIs look different GPUs from the system), and a CUDA kernel can be launched on one of the CI(s) by designating the associated UUID to an environment variable (CUDA\_VISIBLE\_DEVICES), \textit{which is a convenient feature for a job scheduler to handle the co-location and the partitioning in a top-down manner.} 
The partitioning granularity is at GPC (Graphics Processing Cluster) that consists of multiple SMs (Streaming Multiprocessors), and one SM contains FPUs(32/64bit), ALUs, Tensor Cores, a register file, etc. Also, one of the GPCs must be turned off when the MIG feature is enabled for A100. 

By using this hierarchical partitioning feature, we can divide a GPU in a different way regarding how we allocate memory resources to the partitions. 
This is based on the following characteristics: (1) on one hand, all the CIs launched on a GI share lower-level memory hierarchy resources, i.e., Last Level Caches (LLCs), memory controllers, and HBM main memory stacks, inside of the GI; (2) on the other hand, these memory resources are completely partitioned between different GIs. 
These characteristics offer us two partitioning options, as shown in the figures: \textit{private} or \textit{shared} LLC/HBM between CIs. 
\textit{The private option is preferred when (1) the interference between co-scheduled applications is significant, but (2) they both don't require high memory bandwidth. 
The shared option is useful to fully utilize the memory resources especially when one of the co-located applications is memory intensive so that it can utilize the available chip bandwidth as much as possible. 
}

\section{Observations}\label{motivation}

To gain some insights to optimize the resource partitioning, job allocations, and power capping on an MIG-enabled GPU, we perform several preliminary evaluations in this section. 
As scalability is a widely utilized concept/metric to determine the resource allocations~\cite{scalability}, we first observe the scalability of various HPC kernels while changing the GPC allocation. 
We then observe the co-scheduling throughput by co-running multiple applications while changing the resource partitioning and job allocations.

In the scalability evaluation, we launch only one application and test two different memory resource sharing options: private and shared. 
For the private partitioning option, when we scale the number of GPCs for a given application, the number of LLC/HBM modules allocated to it is also increased (and thus the available memory bandwidth scales as well). 
More specifically, when we utilize 1, 2, 3, 4, or 7 GPCs with the private option, 1, 2, 4, 4, or 8 LLC/HBM modules are assigned to the workload respectively, i.e., we scale the size of a GI and allocate the same amount of GPCs to a CI inside of the GI. 
Note we cannot allocate 5 or 6 GPCs to a CI/GI as it is not supported in the current version of the MIG feature.  
As for the shared option, we first create one GI that occupies the entire GPU and then scale the size of CI inside of it. Therefore, regardless of the scale, the CI can utilize the entire GPU memory resources with this option. 
For the co-scheduling throughput observations, we partition 7GPCs into 4 and 3GPCs with the private or the shared LLC/HBM options shown in Figure~\ref{mig-private}/\ref{mig-shared}. 

As for the applications, we picked up one from each category listed in Table~\ref{classification} in Section~\ref{evaluation}, i.e., TI (Tensor core Intensive), CI (Compute Intensive), MI (Memory Intensive), and US (Un-Scalable), to demonstrate the typical scalability behaviors of the categories. 
To demonstrate the impact of resource partitioning and job allocation decisions on the co-scheduling throughput, we pick up two from the co-run combinations shown in Table~\ref{workload-def} in Section~\ref{evaluation}.  
Note that the details of the evaluation environment will be described in Section~\ref{evaluation}. 

\begin{figure}[t]
  \begin{center}
    \includegraphics[width=0.75\linewidth]{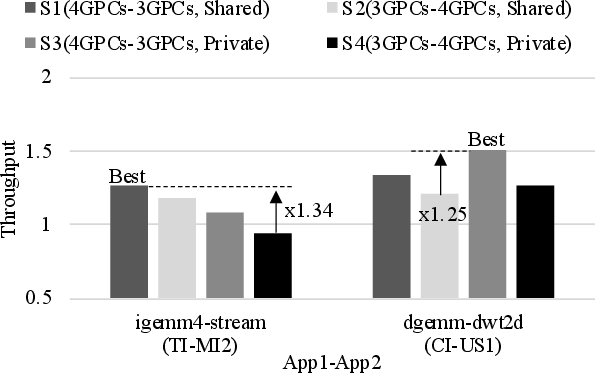}
    \vspace{-5pt}
    \caption{Impact of Resource Partitioning/Allocations on Co-scheduling Throughput (Power Cap: 250[W])}
  \label{corun-motivation}
  \end{center}
  \vspace{-10pt}
\end{figure}

\subsection{Scalability Observations}

Figure~\ref{scale-opt} compares the scalability between the two different partitioning options across several HPC kernels. 
The X-axis indicates the number of GPCs, while the Y-axis represents the relative performance normalized to running without the MIG feature (and thus 8GPCs are available). The power cap is set at 250W in this experiment. As shown in the figure, the scalability and the impact of the mode selection differ depending on the characteristics of the running application. 
\texttt{kmeans} does not fully utilize both the compute and memory resources in this evaluation, and thus scaling down the partition size doesn't affect performance regardless of the option selection. 
\texttt{stream} is a memory intensive workload and thus the selection of the partitioning option matters, while it does not for the other two compute intensive workloads (\texttt{dgemm}/\texttt{hgemm}). 

We then scale down the power cap from 250W to 150W for the shared partitioning option and report the results in Figure~\ref{scale-pow}. 
On one hand, as both \texttt{kmeans} and \texttt{stream} do not fully utilize the power hungry compute resources, the impact of power capping on performance/scalability is negligible for these workloads. 
On the other hand, for compute intensive workloads, in particular \textit{Tensor Core intensive} workloads (e.g., \texttt{hgemm}), the power capping significantly affects the scalability. 
As the scalability is an important factor when we decide the resource allocations, we need to take these hardware/application characteristics into account. 

\subsection{Co-scheduling Throughput}

Next, Figure~\ref{corun-motivation} demonstrates the impact of the resource partition and job allocation decision on co-scheduling throughput. The X-axis shows the executed co-run workloads (TI-MI2 or CI-US1), while the Y-axis indicates the throughput (weighted speedup as described later). 
The left workload consists of \texttt{igemm4} (App1) and \texttt{stream} (App2), and the right one is a mixture of \texttt{dgemm} (App1) and \texttt{dwt2d} (App2). The details will be described in the evaluation section. 
S1-S4 represent resource partitioning and allocation states, and for instance, 4GPCs-3GPCs means that 4GPCs/3GPCs are allocated to App1/App2. 

As shown in the figure, \textit{the best partitioning/allocation choice highly depends on the executed workloads (and their characteristics), and the throughput differs considerably among the selections.} 
TI-MI2 is a mixture of Tensor- and memory-intensive kernels, and thus the following approach works the best: (1) allocating more GPCs to \texttt{igemm4} (Tensor intensive one); and (2) using the shared option in order to allocate more bandwidth for \texttt{stream}.
In the figure, S1 works the best while outperforming the worst one by 34\%.
On the other hand, CI-US1 combines a compute-intensive kernel and an unscalable one, thus the private option performs better. More specifically, this is because: (1) these kernels do not require high memory bandwidth, thus the shared option does not help; and (2) the private option can completely mitigate the interference between them (e.g., shared cache contentions).
For these reasons, S3 works the best for this workload and improves performance by 25\% compared with the worst one. 

These results motivate our methodology to optimize the resource partitioning/allocations as well as the power capping on this new hardware feature, while taking various application characteristics into account. The next section explains the details of our approach. 


\section{Optimizations}\label{proposal}
Motivated by the observations demonstrated in the last section, we propose a methodology to optimize the hardware resource partitioning and application mapping, while considering the power allocation for modern GPUs. 
In this section, we first introduce our workflow to optimize them. 
We second define the problems we solve in this study. 
Third, we provide a linear-regression performance modeling to solve the problems.

\subsection{Workflow Overview}

\begin{figure}[t]
  \begin{center}
    \includegraphics[width=0.9\linewidth]{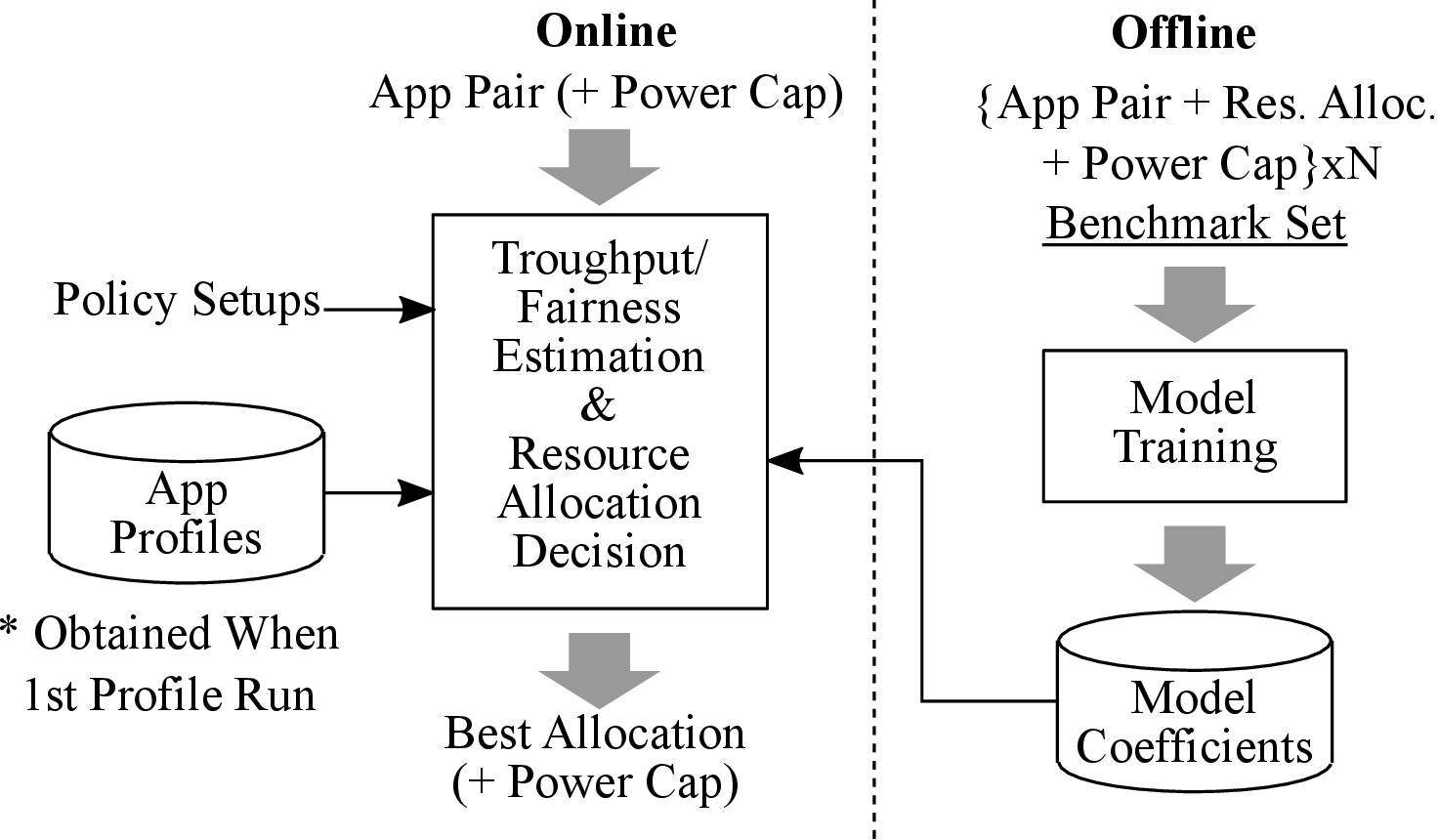}
    \caption{Workflow of Our Approach}
  \label{workflow}
  \end{center}
  \vspace{-10pt}
\end{figure}

Figure \ref{workflow} depicts the overall workflow of our approach that consists of an offline part (left) and an online part (right). 
For the offline part, we train the model coefficients by using predetermined benchmark set. 
As described later, we utilize a linear regression modeling and determine the coefficients by the well-known least square method. 
As for the online part, it solves a given optimization problem (described as the policy setups in the figure) with respect to the resource allocations. 
In this study, we consider two different optimization problems, and depending on the problem (or policy) setup, we optimize the power cap in addition to the resource allocations, as described later. 
The model needs the application characteristics, which is given by the associated profile --- thus profiling is needed for the first run. 
Therefore, if no profile is recorded for an application, then that can not be a target of the co-scheduling and hence must be executed exclusively for the profile run. 


\subsection{Problem Setups and Formulations}
Next, we define and formulate the optimization problems (or policies) to solve in this study. First, the most basic one is described as follows (referred to as \textbf{Problem1}): 
\begin{eqnarray}
&given& App1, App2, \cdots, P, \alpha\nonumber\\
&\max& Throughput(App1, App2, \cdots, S, P)\nonumber\\
&s.t.& Fairness(App1, App2, \cdots, S, P) > \alpha\nonumber\\
&output& S\nonumber
\end{eqnarray}
$App1, App2, \cdots$ are applications to be co-scheduled on the GPU. 
Note that we describe the formulations in this general way, although we co-schedule up to two applications in the experiment. 
$P$ is the power capping value to be set on the GPU, which is a given parameter for this problem. 
$S$ is the resource partitioning state that describes the GPC allocations to the applications and the partitioning option (shared or private). 
In this optimization problem, we try to maximize throughput ($Throughput$) under a fairness constraint ($Fairness > \alpha$) by optimizing the resource partitioning state $S$. 

Next, we solve also the following problem for the further sophisitication (referred to as \textbf{Problem2}): 
\begin{eqnarray}
&given& App1, App2, \cdots, \alpha\nonumber\\
&\max& Throughput(App1, App2, \cdots, S, P) / P \nonumber\\
&s.t.& Fairness(App1, App2, \cdots, S, P) > \alpha\nonumber\\
&output& S, P \nonumber
\end{eqnarray}
In this problem, we optimize the power cap setup (not a given parameter) as well as the resource allocations. 
Here, we also change the objective function to the throughput divided by the power cap --- namely, we consider energy efficiency here. 
As introduced in Section~\ref{background}, the job scheduler needs to determine the power budget allocations aside from the resource partitioning/allocations, and thus optimizing them at the same time would be ultimately an important addition to the job management system. 


For those optimization problems, we need to define both metrics for $Throughput$ and $Fairness$. 
As for the $Throughput$, we utilize the widely-utilized weighted speedup metric~\cite{metrics, metric2} as follows: 
\begin{eqnarray}
& &Throughput(App1, App2, \cdots, S, P) \nonumber\\
&=& WeightedSpeedup(App1, App2, \cdots, S, P) \nonumber\\
&=& \sum_{i} RPerf_{Appi}(S,P)\nonumber
\end{eqnarray}
Here, $RPerf_{Appi}(S,P)$ is relative performance of $Appi$ when the hardware setups are $S, P$, which is normalized to performance of $Appi$ when exclusive solo run without power capping nor resource partitioning. If the weighted speedup is greater than $1$, then it is better than the time-sharing executions in chip throughput. 

For the $Fairness$, we simply define as follows: 
\begin{eqnarray}
& &Fairness(App1, App2, \cdots, S, P)\nonumber\\ 
&=& \min(RPerf_{App1}(S,P), RPerf_{App2}(S,P), \cdots)\nonumber
\end{eqnarray}
We choose this simple minimum function for the fairness metric aiming to avoid the situation where an application suffers from a significant slowdown by co-scheduling or power capping. 


For both metrics, we need to estimate the relative performance $RPerf_{Appi}(S,P)$ for all the co-located applications. 
For this, we construct a simple performance model, which we introduce in the next subsection. 
Note the parameters described in this subsection (and also the next subsection) are listed in Table~\ref{param-def}.


\begin{table}[b]
{
\scriptsize
\caption{Definitions of Parameters/Function}\label{param-def}
\vspace{-10pt}
\begin{center}
\begin{tabular}{|M{0.25\linewidth}||M{0.60\linewidth}|}
\hline
Symbol&Explanation\\\hline\hline
$P$ & The power cap set to the entire GPU chip\\\hline
$S$ & The state of hardware partitioning and GPC allocations to the co-located applications\\\hline
$\alpha$ & The threshold parameter to assure the fairness\\\hline
\hline
$Appi$ & ith application to be co-located on the GPU chip\\\hline
$\mathbf{F_{Appi}}$ & The features of $Appi$, i.e., the vector of performance counter values for $Appi$\\\hline
\hline
$RPerf_{Appi}(S,P)$ & Relative performance of $Appi$ when the hardware state is $(P, S)$, normalized to that of exclusive solo run without partitioning nor power capping\\

\hline

\end{tabular}
\end{center}
}
\end{table}

\subsection{Linear Regression Performance Modeling}

To estimate $RPerf_{Appi}(S,P)$, the relative performance of $Appi$ as a function of $S$ and $P$, we apply the well-known linear regression performance modeling~\cite{linear-regression, lr-application1, lr-application2, lr-application3, lr-application4, lr-application5}. 
More specifically, we model $RPerf_{Appi}(S,P)$ as follows: 
\begin{eqnarray}
RPerf_{Appi}(S,P) = \mathbf{C}(S, P)\cdot\mathbf{H}(\mathbf{F_{Appi}}) + \sum_{j\neq i} \mathbf{D}(S, P)\cdot\mathbf{J}(\mathbf{F_{Appj}})\nonumber
\end{eqnarray}
The first term $\mathbf{C}(S, P)\cdot\mathbf{H}(\mathbf{F_{Appi}})$ represents the scalability of this application, i.e., how much performance improvement can be achieved by scaling the number of GPCs and the power cap value. 
This highly depends on the application characteristics, and thus here we consider application features $\mathbf{F_{Appi}}$, i.e., the vector that holds performance counter values of $App_i$ obtained when the profile run, which is ultimately converted into another vector by a basis function $\mathbf{H}$. 
The second term represents the interference impact from the other co-located applications on performance of $Appi$. 
This term also utilizes the application features $\mathbf{F_{Appj}}$, which is also converted into another vector by a different basis function $\mathbf{J}$. 
For both of the terms, the coefficients vectors ($\mathbf{C}$, $\mathbf{D}$) are functions of the hardware state ($S$, $P$). 
This means we apply the least square method (or curve fitting) for each combination of ($S$, $P$) independently and separately to estimate those coefficients.  
In our evaluation, the number of combination is very limited as described later, and thus this is not an issue.  
Moreover, even if the number would explode for future hardware, several existing methodologies could be applicable to estimate them efficiently~\cite{lr-application5}.  
Further, we can ignore the second term for the training/inference when we execute only one application but change the hardware resource/power allocations.

\begin{figure*}[t]
  \begin{center}
    \includegraphics[width=\linewidth]{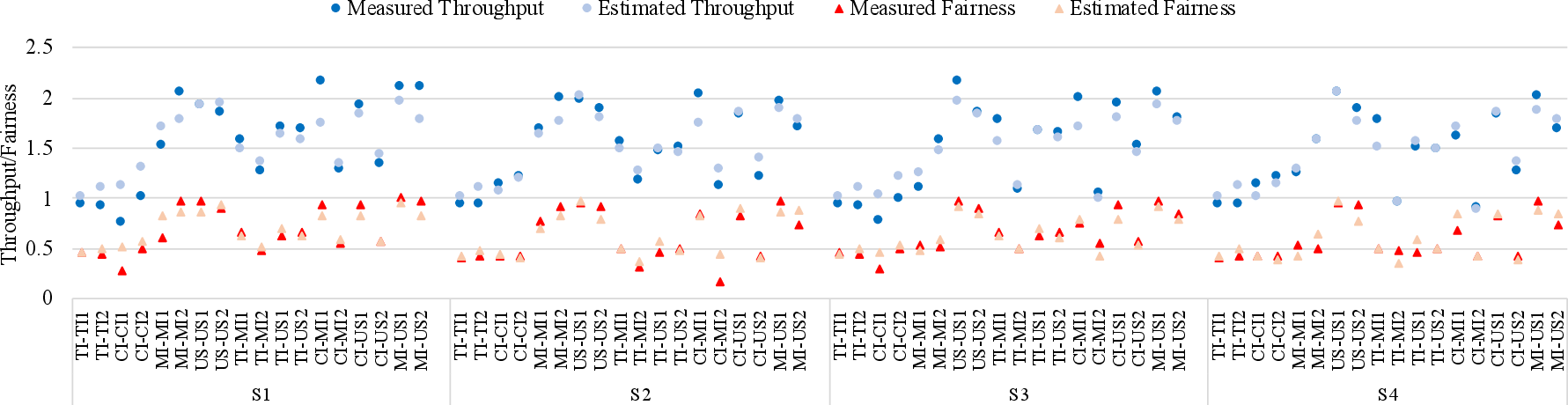}
    \caption{Comparison of Estimated and Measured Throughput/Fairness across Different Workloads (P=250W)}
  \label{error-250w}
  \end{center}
\end{figure*}

\section{Evaluation}\label{evaluation}
In this section, we first describe our evaluation environment and methodology in Section~\ref{environment}. 
We then provide and analyze our experimental results in Section~\ref{experiment}.

\subsection{Evaluation Setup}\label{environment}

\subsubsection{Evaluation Environment}
In this evaluation, we use the platform summarized in Table~\ref{time-x-system}. 
Our approach is applicable if the GPU in a target system supports both a resource partitioning and a power capping features. 
For this, we utilize an A100 card as it meets the requirement while offering the MIG capability~\cite{a100}.
To collect performance counter values during each profile run, we utilize a profiling framework for NVIDIA GPUs, called NSight Compute~\cite{nsight-compute}. 
By using the profiler, we collect the performance counter values listed in Table~\ref{counters}. 
The definitions of these statistics here are provided in the tool. Table~\ref{functions} lists the basis function setups, which are based on the evaluations demonstrated in Section~\ref{motivation}. 

Table~\ref{search-space} summarizes the partitioning and power capping settings we explore in this evaluation. We examine both the private and the shared LLC/HBM options, and the GPU is divided into 4GPCs and 3GPCs for both of the options. Although the current architecture does not support a flexible partitioning option (e.g., no partitioning support for 6GPCs/1GPC nor 5GPCs/2GPCs), our approach is extensible to cover it for future GPUs. As for the power capping, we scale it from 150W to 250W at the granularity of chip by using the nvidia-smi command. Finer-grained power capping, such as at GPC level, would be useful, and our approach could cover this option as well with a minor fix in the model coefficients if it would be supported in future GPUs.

\begin{table}[b]
{
\scriptsize
\caption{System Configurations}\label{time-x-system}
\vspace{-10pt}
\begin{center}
\begin{tabular}{ |M{0.25\linewidth}||M{0.60\linewidth}| } 
\hline
 Name & Remarks \\\hline\hline
 GPU &  NVIDIA A100 40GB PCIe 4.0 \\\hline
 CPU \& Memory & AMD Ryzen Threadripper PRO 3955WX 16-Cores, DIMM DDR4 3200 MHz 32GB (Total) 2ch \\\hline
 Operating System & Ubuntu 20.04, Kernel Version: 5.15.5 \\\hline
 Compiler \& Driver & CUDA Version: 11.5, Driver Version: 495.29.05 \\\hline
\end{tabular}
\end{center}
}
\end{table}

\begin{table}[b]
{
\scriptsize
\caption{Collected Performance Counters ($\mathbf{F}$)}\label{counters}
\vspace{-10pt}
\begin{center}
\begin{tabular}{ |M{0.95\linewidth}| } 
\hline
Counters and Definitions
\\\hline\hline
$F_1$ = Compute Throughput \lbrack\%\rbrack, $F_2$ = Memory Throughput \lbrack\%\rbrack, $F_3$ = DRAM Throughput \lbrack\%\rbrack, $F_4$ = L2 Hit Rate \lbrack\%\rbrack, $F_5$ = Occupancy \lbrack\%\rbrack, $F_6$ = Tensor (MIXED) \lbrack\%\rbrack, $F_7$ = Tensor (DOUBLE) \lbrack\%\rbrack, $F_8$ = Tensor (INTEGER) \lbrack\%\rbrack
\\\hline 
\end{tabular}
\end{center}
}
\end{table}
\begin{table}[b]
{
\scriptsize
\caption{Basis Function Setups ($\mathbf{H(F)}$, $\mathbf{J(F)}$)}\label{functions}
\vspace{-10pt}
\begin{center}
\begin{tabular}{ |M{0.95\linewidth}| } 
\hline
Function Definitions 
\\\hline\hline
$H_{1}=F_{1}/100 - H_{2}$ (non tensor compute intensity), $H{2}=(F_{6}+F_{7}+F_{8})/100$ (tensor compute intensity), $H{3}=F_{2}/F_{1}$ (memory/compute ratio), $H{4}=F_{4}/100$ (DRAM intensity), $H_{5}=F_{5}/100$ (resource utilization), $H_{6}=const.$\\\hline
$J_{1}=F_{3}/100$ (DRAM intensity), $J_{2}=F_{4}/100$ (access pattern related), $J_{3}=const.$
\\\hline 
\end{tabular}
\end{center}
}
\end{table}

\begin{table}[b]
{
\scriptsize
\caption{Power Cap and Partitioning Selections}\label{search-space}
\vspace{-10pt}
\begin{center}
\begin{tabular}{ |M{0.25\linewidth}||M{0.60\linewidth}| } 
\hline
 Variable & Selections \\\hline\hline
 $P$ &  150, 170, 190, 210, 230 250 [W] \\\hline
 $S$: (GPCs for App1, GPCs for App2, LLC/HBM option) & S1=(4GPCs, 3GPCs, Shared), S2=(3GPCs, 4GPCs, Shared), S3=(4GPCs, 3GPCs, Private), S4=(3GPCs, 4GPCs, Private)\\\hline
\end{tabular}
\end{center}
}
\end{table}

\subsubsection{Workloads}
To evaluate our approach, we utilize the Rodinia benchmark suite because it is widely used for a variety of studies on heterogeneous systems~\cite{rodiana}. 
However, as Rodinia benchmarks hardly utilize Tensor Cores 
thus we also include the Cutlass library (in particular we use Cutlass Profiler) that offers Tensor Core intensive kernels~\cite{cutlass}. 
More specifically, we evaluate various DGEMM kernels with different Tensor operations, details of which are summarized in Table~\ref{cutlas-options}. 
Further, we also utilize several basic benchmarks including \texttt{stream}~\cite{stream} and \texttt{randomaccess}~\cite{random}. 

For the co-scheduling evaluation, we provide co-run pairs in the following way. 
First, we classify these benchmarks into \textit{Tensor Core Intensive (TI)}, \textit{Non Tensor Compute Intensive (CI)}, \textit{Memory Intensive (MI)}, or \textit{Un-Scalable (US)} based on the scalability observations as well as the collected performance counter statistics in the following way. 
First, if the performance degradation at 150W with 1GPC using the private option is less than 10\%, it is classified into \textit{US}. 
Otherwise, we check $F_1$/$F_2$ and if it is greater than 0.80, then the benchmark is categorized as \textit{TI} or \textit{CI} (\textit{TI} if it utilizes the Tensor Cores) otherwise it belongs to $MI$. 
The classification is listed in Table~\ref{classification}. 
Next, we create pairs of classes, such as \textit{TI-MI} and \textit{CI-CI}, and then select an application for each class in each pair. 
The application is chosen randomly from the associated class. 

\begin{table}[b]
{
\scriptsize
\caption{Workload Specifications for DGEMM Variants}\label{cutlas-options}
\vspace{-10pt}
\begin{center}
\begin{tabular}{ |M{0.20\linewidth}||M{0.65\linewidth}| } 
\hline
 Name & Specifications \\\hline\hline
 sgemm & Normal SGEMM without using Tensor Cores \\
 dgemm & Normal DGEMM without using Tensor Cores\\
 tdgemm & DGEMM with Tensor Cores \\
 tf32gemm & GEMM using TF32 for inputs and FP32 for accumulation \\
 hgemm & HGEMM using FP16 for both inputs and accumulation \\
 fp16gemm & GEMM using FP16 for inputs and FP32 for accumulation \\
 bf16gemm & GEMM using BF16 for inputs and FP32 for accumulation \\
 igemm4 & IGEMM using u4 for both inputs and accumulation \\
 igemm8 & IGEMM using u8 for both inputs and accumulation \\
 \hline
\end{tabular}
\end{center}
} 

{
\scriptsize
\caption{Benchmark Classifications}\label{classification}
\vspace{-10pt}
\begin{center}
\begin{tabular}{ |M{0.10\linewidth}||M{0.75\linewidth}| } 
\hline
 Class & Benchmarks \\\hline\hline
 TI & tdgemm, tf32gemm, hgemm, fp16gemm, bf16gemm, igemm4, igemm8 \\\hline
 CI & hotspot, lavaMD, sgemm, dgemm, srad, heartwell \\\hline
 MI & randomaccess, stream, gaussian, leukocyte, lud  \\\hline
 US & backprop, bfs, dwt2d, kmeans, needle, pathfinder \\
\hline
\end{tabular}
\end{center}
}
{
\scriptsize
\caption{Co-run Workloads Definitions}\label{workload-def}
\vspace{-10pt}
\begin{center}
\begin{tabular}{ |M{0.95\linewidth}| } 
\hline
Name = (App1, App2)
\\\hline\hline
TI-TI1 = (tdgemm, tr32gemm), TI-TI2 = (fp16gemm, bf16gemm),  CI-CI1 = (sgemm, lavaMD), 
CI-CI2 = (dgemm, hotspot), 
MI-MI1 = (randomaccess, gaussian), 
MI-MI2 = (stream, leukocyte), 
US-US1 = (bfs, dwt2d), 
US-US2 = (kmeans, needle), 
TI-MI1 = (hgemm, lud), 
TI-MI2 = (igemm4, stream), 
CI-MI1 = (heartwell, gaussian),
CI-MI2 = (sgemm, randomaccess),
TI-US1 = (igemm8, backprop), 
TI-US2 = (fp16gemm, pathfinder),
CI-US1 = (srad, needle),
CI-US2 = (dgemm, dwt2d),
MI-US1 = (leukocyte, kmeans),
MI-US2 = (lud, needle)
\\\hline 
\end{tabular}
\end{center}
}
\end{table}

\subsubsection{Methodology}
We first collect a profile for each benchmark while executing it without any power capping, partitioning or co-scheduling. 
We then execute these benchmarks with exclusive solo-runs while scaling the power and hardware resource allocations (i.e., the number of GPC allocations with different partitioning options) and measure performance degradation. Then, the solo-run coefficients $\mathbf{C}$ are estimated by applying the linear regression to the combinations of the performance degradation and the profiles (obtained in the first step) for each hardware setup. 
We next execute co-run workloads while changing the power and hardware resource allocations, measure performance degradation as well, and then estimate $\mathbf{D}$ by using the profiles of co-located applications, along with the measured performance. 
We finally confirm the accuracy of our model and check if we can choose an optimal combination of power cap, partitioning, application mappings for a given problem. 
To pick up the optimal combination, we simply utilize the exhaustive search because the number of selections here is very small in this study ($4\times6=24$). 
In case the search space would increase significantly in the future work (e.g., by covering much more hardware setup options), we could simply apply some heuristics here such as the hill-climbing algorithm, following prior studies~\cite{lr-application5}. 
Throughout the evaluation, we normalize performance to that without both MIG and power capping.

\begin{figure}[t]
  \begin{center}
    \includegraphics[width=\linewidth]{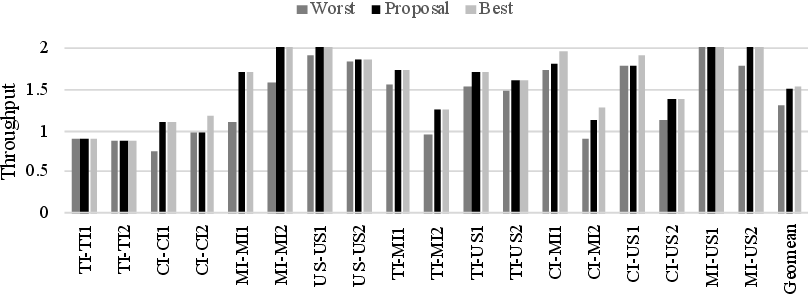}
    \vspace{-15pt}
    \caption{Throughput Comparison at P=230W and $\alpha$ = 0.2}
  \label{proposal-230w}
  \end{center}
  \begin{center}
    \includegraphics[width=0.6\linewidth]{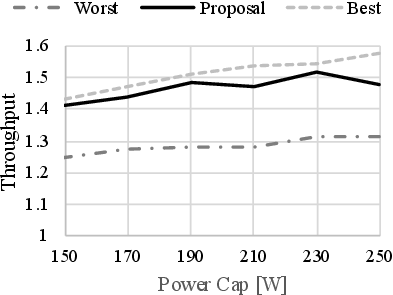}
    \vspace{-5pt}
    \caption{Throughput (Geometric Mean) as a Function of Power Cap ($\alpha$ = 0.2)}
  \label{geomean}
  \end{center}
  \vspace{-10pt}
\end{figure}

\begin{figure*}[t]
{
\begin{tabular}{c}

 \begin{minipage}{\hsize}
  \begin{center}
  \begin{minipage}{0.48\hsize}
  \begin{center}
  \includegraphics[width=\linewidth]{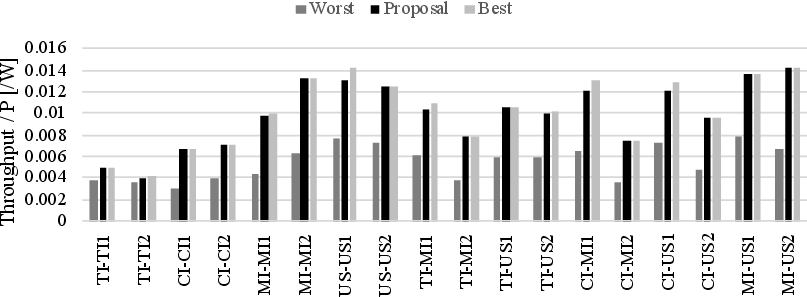}
  \textbf{$\alpha = 0.20$}
  \end{center}
  \end{minipage}
  \hspace{0.01\hsize}
  \begin{minipage}{0.48\hsize}
  \begin{center}
  \includegraphics[width=\linewidth]{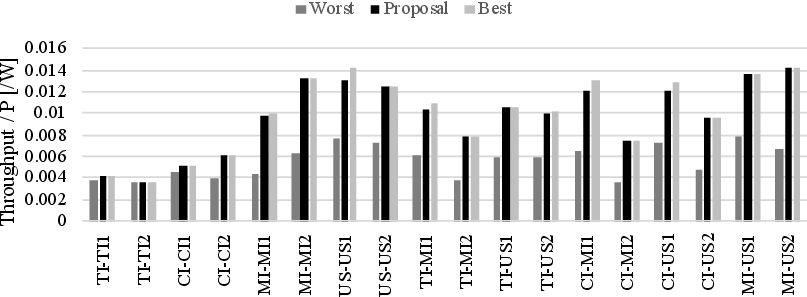}
  \textbf{$\alpha = 0.42$}
  \end{center}
  \end{minipage}
  \end{center}
  \end{minipage}
\end{tabular}
  \vspace{-5pt}
    \caption{Energy Efficiency (Throughput / P) Comparison across Different Workloads}\label{energy-efficiency}
\vspace{5pt}
\begin{tabular}{c}

 \begin{minipage}{\hsize}
  \begin{center}
  \begin{minipage}{0.48\hsize}
  \begin{center}
  \includegraphics[width=\linewidth]{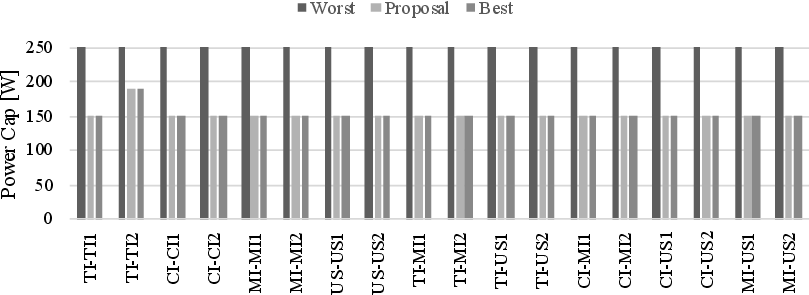}
  \textbf{$\alpha = 0.20$}
  \end{center}
  \end{minipage}
  \hspace{0.01\hsize}
  \begin{minipage}{0.48\hsize}
  \begin{center}
  \includegraphics[width=\linewidth]{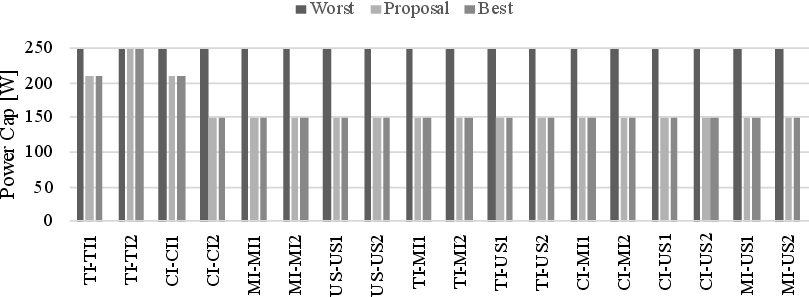}
  \textbf{$\alpha = 0.42$}
  \end{center}
  \end{minipage}
  \end{center}
  \end{minipage}
\end{tabular}
  \vspace{-5pt}
    \caption{Power Budgeting Comparison across Different Workloads}\label{power}
}
\end{figure*}

\subsection{Experimental Result}\label{experiment}
\subsubsection{Model Accuracy}
Figure~\ref{error-250w} compares the measured and estimated throughput/fairness metrics for different workloads with different partitioning and allocation states at $P$=250W. 
The horizontal axis lists different co-run workloads with different partitioning/allocation status ($S$), while the vertical axis indicates the throughput or fairness. 
As shown in the figure, our linear-regression modeling captures the characteristics of both the hardware and the running workloads and convert them into these performance metrics well. 
We observed the similar accuracy trends for the other power cap setups from 150W to 230W, and the average error (i.e., average of absolute differences divided by the measured value) across all the workloads and hardware setups is around 9.7\% for the throughput metric and 14.5\% for the fairness metric. 
As shown later, these errors do not have any significant impacts on the decision making for these optimization problems.

\subsubsection{Problem1}

We solve the first optimization problem by using the performance model. Figure~\ref{proposal-230w} illustrates the evaluation result when the power cap and $\alpha$ are set at 230W and 0.2, respectively. 
The X-axis lists the evaluated workloads, while the Y-axis indicates the throughput. 
The worst/best chooses one partitioning/allocation state (S) from those meet the fairness constraint so that the throughput is minimized/maximized.  
In the figure, \textit{Geomean} is the geometric mean of throughput across these workloads for the best and worst case and for the proposal.  
Overall, our approach successfully selects the right partitioning/allocation setup state (S), and the throughput is close to the best in the geometric mean (1.52/1.54 for proposal/best).

Next we scale the power cap allocation from 150W to 250W and compare the geometric mean of throughput among the worst, proposal, and best. 
Figure~\ref{geomean} presents the result. The X-axis shows the allocated power cap, while the Y-axis indicates the throughput. 
As shown in the figure, our approach is successful and is close to the best setup for most of the power cap setup. 
Note that no fairness violation happened for our approach in this evaluation.

\begin{figure}[t]
  \begin{center}
    \includegraphics[width=0.6\linewidth]{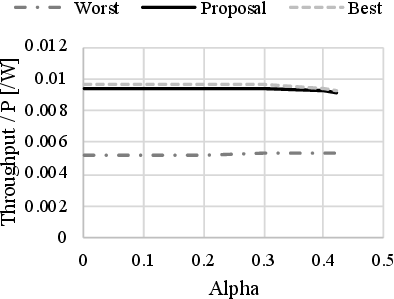}
    \caption{Geometric Mean of Energy Efficiency as a Function of $\alpha$}
  \label{enaeff-alpha}
  \end{center}
  \vspace{-15pt}
\end{figure}

\subsubsection{Problem2}
We attempt to solve the last problem that targets the optimization of the power cap and resource partitioning/allocation setups at the same time to maximize the energy efficiency metric, i.e., $Throughput/P$. 
We solve this problem while changing the Fairness requirement parameter from $\alpha$ from 0.2 to 0.42. 
Note we picked up 0.42 as the maximum threshold so that at least one or more solution(s) exist(s) for all the workloads --- this is not the case when $\alpha \geq 0.43$. 
Figure~\ref{energy-efficiency} shows the energy efficiency comparison among the worst, our proposal, and the best case. 
Here, the worst/best case chooses the combination of the power cap and the hardware state (P, S) from those meeting the fairness constraint so that the energy efficiency metric is minimized/maximized. 
As shown in the figures, our approach reaches almost the best energy efficiency regardless of the given workloads or the fairness parameter setup. 

Next, we demonstrate the comparison of power cap selections among the worst, our proposal, and the best case across different workloads in Figure~\ref{power}. 
Our approach can correctly set the power cap value for these workloads and, at the same time, is sensitive to the fairness parameter $\alpha$. 
By correctly setting the power cap to given workloads, we can improve the total HPC system throughput or energy efficiency by shifting the extra power budget to where it can be used more efficiently (e.g., to a compute-intensive node with requiring high performance/fairness). 
Note our approach did not cause any fairness violations in this evaluation as well.

Finally, we present the energy efficiency as a function of the threshold parameter ($\alpha$) in Figure~\ref{enaeff-alpha}. 
The horizontal axis indicates the threshold, while the vertical axis shows the geometric mean of the energy efficiency across all the workloads. 
As shown in the figure, our approach is very close to the best setup regardless of the threshold parameter setting. 
Note that no fairness violation happened for our approach in this evaluation as well. 
\section{Discussions}\label{discussions}
As the GPU partitioning features has become available only very recently, the applicability of our approach is currently limited to some commercial GPUs, i.e., some high-end Ampere generation NVidia GPUs. 
However, given the situation that the improvement of resource utilization is becoming more and more important in many large-scale systems, including supercomputers and datacenters, GPU partitioning features will be widely supported in the future. 
Our approach is applicable only if the target GPU supports a resource partitioning mechanism at compute resources with/without different memory sharing options. 
Even if future GPUs would support more flexible and finer-grained partitioning options, needed modifications to our approach would be trivial: (1) listing all the possible power allocation states for $P$ and all the possible partitioning and job allocations states for $S$; (2) applying the linear regression for all the possible combinations ($P$, $S$). 
One limitation of our approach is that we manually set the selections of hardware performance counters and basis functions. However, this part can be automated by importing methodologies from some other studies in the literature~\cite{perf-counter1, perf-counter2}. Another limitation is the selection of the best setups and the calibration of coefficients would be time consuming if the numbers of $P$ and $S$ would increase because we apply the exhaustive search. 
However, we can also import some methodologies here from other studies as well~\cite{lr-application5}.

\section{Related Work}\label{related}
In this section, we introduce the literature of co-scheduling (or multi-processing) and performance modeling, and then we clarify the uniqueness of this study compared with the prior studies. 

\subsection{Multi-processing Support on GPUs}

S. Pai et al. point out the waste of resources within a GPU when running a CUDA kernel and explored the feasibility of multi-processing using their elastic kernel implementation~\cite{elastic-kernel}. 
I. Tanasic et al. propose a microarchitectural mechanism to enable multi-processing on GPUs without any kernel modifications~\cite{preemptive-gpu}. 
As a consequence, a similar mechanism called MPS has been already supported in commercial NVIDIA GPUs~\cite{mps}. 

Several studies propose software mechanisms to improve the efficiency of multi-processing. 
T. Allen et al. propose a framework called Slate that optimizes the combination of co-located processes and dynamically adjusts the scales of them~\cite{slate}. 
smCompactor is a similar framework to Slate, while its major focus is on maximizing the resource utilization~\cite{smcompactor}. 
C. Reano et al. propose a safe co-scheduling mechanism that takes memory footprints into account when processes are co-scheduled in a time sharing manner~\cite{time-sharing}. 
Although these techniques (including MPS) are useful to improve the job throughput, they are node-level runtime process schedulers that independently optimize process/resource allocations at each node, and are less suitable for multi-node HPC jobs. 
\textit{In contrast, hardware-level partitioning features, like MIG~\cite{mig}, divide a GPU into multiple parts, which are recognized as multiple different GPUs with different UUIDs by the system, making the job manager easier to co-locate single-/multe-node jobs, and thus we are focusing on this feature for the future extensions/integration. }

There are several microarchitectural techniques to improve the efficiency of the concurrency controlling features~\cite{hw-space-sharing, hw-space-sharing2}. They are also orthogonal to our study, and  they would ultimately improve the efficiency of GPUs.

\subsection{Power Capping and Co-scheduling}
Ever since multi-core processors appeared on the market, the combination of power capping and co-scheduling have been widely studied. 
R. Cochran et al. propose Pack \& Cap that co-locates multi-threaded applications on a multi/many-core processor while optimizing the number of threads for each of the applications~\cite{cpu-cosh-powcap}. 
H. Sasaki et al. extend the idea by targeting also power saving of idle cores~\cite{cpu-cosh-powcap2}. 
Q. Zhu et al. work on co-scheduling CPU-only and GPU jobs on a CPU-GPU heterogeneous system under power capping~\cite{cpu-gpu-cosh}. 
K. Straube et al. propose an efficient power capping mechanism for GPUs~\cite{gpu-pow-cap}, which can be ultimately applied for co-scheduled environments. 
The major uniqueness of this study compared with them is that only ours considers the combination of power capping, coarse-grained co-scheduling (i.e., MIG), and intra-node heterogeneity (i.e., Tensor Cores) for modern GPUs. 

\subsection{Statistical Performance Modeling}
There have been a variety of hardware performance counter based performance modeling studies using statistical approaches~\cite{linear-regression, lr-application1, lr-application2, lr-application3, lr-application4, lr-application5}. 
B. Lee et al. apply linear regression modeling to performance estimation for CPUs~\cite{linear-regression}. 
H. Sasaki et al. extend the approach to predict the impact of DVFS (Dynamic Voltage and Frequency Scaling) on performance to choose the best clock frequency setup~\cite{lr-application1}. 
B. Barnes et al. propose a statistical approach to predict performance of parallel applications~\cite{lr-application2}. 
H. Nagasaka et al. model power consumption of GPUs by using hardware performance counters and statistical modeling~\cite{lr-application3}. 
Similar approaches have been also utilized for more complicated system optimization purposes, such as clock frequency setups on both CPU and GPU at the same time~\cite{lr-application4} and power capping setups on CPU, DRAM, and NVRAM~\cite{lr-application5}. 
We follow the literature and tailored the linear regression to apply it to throughput and fairness prediction on a modern GPU partitioning mechanism under power caps.

\section{Conclusion}\label{conclusion}
In this paper, we targeted the combination of GPU partitioning and power capping, provided a systematic methodology to optimize these hardware setups and job allocations, and demonstrated the effectiveness using the MIG feature and power capping mechanism supported in the most recent GPUs. 
First, we scaled the resource allocations, i.e., the number of GPCs and the power limit setup, for various applications or kernels, and observed the impact on performance. As a result, it turned out that the scalability highly depends on the application characteristics such as the compute/memory intensity and the Tensor Core utilization, as well as the power cap setup. 
Second, based on the preliminary evaluation results, we defined our optimization problem, proposed a framework, and introduced a simple performance model, aiming to optimize the resource allocations for co-located applications while taking various application characteristics as well as power managements into account. 
The experimental result indicates that our approach is successful in selecting an optimal combination, achieving almost the best throughput/energy-efficiency across multiple different workloads.

In our future work, we will extend this work to cover the job scheduler side, i.e., selecting an optimal combination of co-locating jobs from a job queue at cluster scale, which will require the integration/interaction with an existing scheduler, such as SLURM. 
Another option is covering other components (e.g., CPUs and memories) and optimizing power budgeting across them when co-locating different kind of jobs including such as CPU-only jobs. 
Furthermore, if future commercial GPUs would support more flexible partitioning or power management features, then we would extend this work accordingly to handle them efficiently.

\balance


\begin{acks}
This work has received funding under the European Commission's EuroHPC and H2020 programmes under grant agreement no. 956560 and was supported by the NVIDIA Academic Hardware Grant Program. In addition, MEGWARE Computer Vertrieb und Service GmbH helped this work with giving us exclusive accesses to their testbeds in the early stage of this work. 
\end{acks}

\bibliographystyle{ACM-Reference-Format}
\bibliography{ref.bib}

\end{document}